\documentclass[aip, preprint, superscriptaddress, amsmath, amssymb]{revtex4-2}

\usepackage{color}
\usepackage{caption}
\usepackage[labelformat=simple]{subcaption}
\usepackage{graphicx}
\usepackage{amssymb}
\usepackage{amsmath}
\usepackage[hidelinks=true]{hyperref}
\hypersetup{
		colorlinks = true,
		urlcolor = blue,
		linkcolor = blue,
		citecolor = blue
	}

\newcommand{\C}{$^\circ$C }    

\begin{document}

\title{Strain-dependent Insulating State and Kondo Effect in Epitaxial SrIrO$_{3}$ Films}

\author{Gaurab Rimal}
\email{gaurab.rimal@wmich.edu}
\affiliation{Department of Physics, Auburn University, Auburn, AL, USA}
\affiliation{Department of Physics, Western Michigan University, Kalamazoo, MI, USA}

\author{Tanzila Tasnim}
\affiliation{Department of Physics, Auburn University, Auburn, AL, USA}

\author{Gabriel Calderon Ortiz}
\affiliation{Department of Materials Science and Engineering, Ohio State University, Columbus, OH, USA}

\author{George E. Sterbinsky}
\affiliation{Advanced Photon Source, Argonne National Laboratory, Lemont, IL, USA}

\author{Jinwoo Hwang}
\affiliation{Department of Materials Science and Engineering, Ohio State University, Columbus, OH, USA}

\author{Ryan B. Comes}
\email{ryan.comes@auburn.edu}
\affiliation{Department of Physics, Auburn University, Auburn, AL, USA}

\newcommand{\red}[1]{\textcolor{red}{#1}}
\newcommand{\blue}[1]{\textcolor{blue}{#1}}
\newcommand{\green}[1]{\textcolor{green}{#1}}

\begin{abstract}
    The large spin-orbit coupling in iridium oxides plays a significant role in driving novel physical behaviors, including emergent phenomena in the films and heterostructures of perovskite and Ruddlesden-Popper iridates. In this work, we study the role of epitaxial strain on the electronic behavior of thin SrIrO$_3$ films. We find that compressive epitaxial strain leads to metallic transport behavior, but a slight tensile strain shows gapped behavior. Temperature-dependent resistivity measurements are used to examine different behaviors in films as a function of strain. We find Kondo contributions to the resistivity, with stronger effects in films that are thinner and under less compressive epitaxial strain. These results show the potential to tune SrIrO$_3$ into Kondo insulating states and open possibilities for a quantum critical point that can be controlled with strain in epitaxial films.  
\end{abstract}

\maketitle

The subtle role of quantum properties emerging from spin, charge, exchange, correlations and topology bring about a plethora of unexpected emergent electronic behavior in quantum materials \cite{Tokura2017}. The large spin-orbit coupling (SOC) energy  in 4d and 5d transition metal oxides (TMO) compete with similar energy scales arising from other electronic degrees of freedom, including crystal-field splitting, magnetic coupling, and electronic correlations, which brings about a rich playground that helps realize new and unexpected electronic phases \cite{Witczak-Krempa2014}. Furthermore, electron confinement and hybridization at the interface of heterostructures leads to emergent electronic states such as electron gases, superconductivity, magnetism and quantum Hall states \cite{Hwang2012,falson_review_2018}.

One interesting material system from the perspective of high spin-orbit coupling is the iridates \cite{cao_challenge_2018,lu_jeff_2020}. These oxides, which have iridium as the main component, exhibit subtle interplay between spin, orbit, charge and crystallographic degrees of freedom, leading to competing electronic phases \cite{rau_spin-orbit_2016}. For example, the Ruddlesden-Popper (RP) series Sr$_{n+1}$Ir$_n$O$_{3n+1}$ comprises of the insulating Sr$_2$IrO$_4$ (n=1), insulating Sr$_3$Ir$_2$O$_7$ (n=2), and the semi-metallic SrIrO$_3$ (n=$\infty$). The commonality of these systems is that they are correlated materials, and due to the large SOC imparted by Ir, are predicted or show phases such as superconductivity \cite{wang_twisted_2011} and quantum paramagnetism \cite{haskel_possible_2020}. Due to the large SOC from Ir, the t$_{2g}$ band effectively splits into J$_{eff}$=1/2 and J$_{eff}$=3/2 angular momentum states, and the Fermi level is positioned such that the ground state is a J$_{eff}$=1/2 paramagnetic semi-metal \cite{cao_challenge_2018}. Furthermore, Dirac nodes near the Fermi level provide an avenue to tune electronic parameters which can result in metal-insulator transitions and novel magnetism \cite{zeb_interplay_2012}. 


The role of dimensionality in this RP series is also notable. As one goes from $n=1$ (Sr$_2$IrO$_4$) to $n=\infty$ (SrIrO$_3$), the magnetic property of the films changes, from antiferromagnetic Mott insulator to a paramagnetic semimetal. However, BaIrO$_3$, a hexagonal perovskite, is an insulator that shows ferromagnetism and charge density wave states \cite{cao_charge_2000}. The sensitivity to crystal symmetry thus drives the metallicity and magnetic behavior of these materials. Clearly, small changes imparted by thermodynamic or material parameters can dramatically change their properties. 


A key challenge in studying novel phases is obtaining high-quality materials that are free from impurities and defects, such that disorder does not break symmetry and phase coherence. Improvements in material synthesis and characterization techniques has allowed the realization of many novel materials. Although bulk single crystals can be grown using traditional solid state synthesis techniques, thin films have the advantage of tuning the dimensionality, as well as the added benefits such as strain manipulation, delta and modulation doping, and Fermi level control via electrostatic gating. Molecular beam epitaxy (MBE) is a preferred method in thin film growth and typically results in films with the best qualities \cite{Schlom2015}. The ability to control multiple degrees of freedom, combined with \textit{in situ} characterization\cite{thapa2021probing} and ultra-high vacuum transfer to other probe stations makes MBE the best technique to grow ultra-high quality materials. Although widely used in the semiconductor industry, it has now been well adapted for growth of oxides, and new improvements \cite{rimal_advances_2024} have allowed the realization of materials in the 4d and 5d blocks of the periodic table that were previously quite difficult, if not impossible, to grow as thin films. 

The major difficulty in MBE-growth of iridates is the low vapor pressure of Ir, which can be circumvented by using traditional electron-beam source \cite{nie_interplay_2015} or a metalorganic source \cite{choudhary_semi-metallic_2022}. Here, we report the growth of SrIrO$_3$ using metalorganic MBE and show how epitaxial strain can be used to tune the electronic behavior of the material. Unlike previous reports, we show that a slight tensile strain leads to insulating behavior, whereas compressive strain leads to a metallic ground state. Furthermore, we observe a Kondo contribution to resistivity which scales with strain. We posit that strain in epitaxial SrIrO$_3$ films has a large role in the electronic behavior through tuning of oxygen vacancies. The observation of Kondo effect that scales with strain hint at the proximity of the $n = \infty$ phase to a magnetically-ordered ground state.

We grew SrIrO$_{3}$ films with varying thickness on LaAlO$_3$ (LAO),  (La$_{0.18}$Sr$_{0.82}$)(Al$_{0.59}$Ta$_{0.41}$)O$_3$ (LSAT), SrTiO$_3$ (STO) and GdScO$_3$ (GSO) substrates using metalorganic MBE following a previously reported procedure \cite{choudhary_semi-metallic_2022}. The growth temperature was 650 \C, as monitored by a thermocouple placed behind the substrate. The oxygen plasma was kept at a pressure of 6$\times 10^{-6}$ Torr and plasma was generated at a power of 300 W. The films were grown at a rate of approximately 1 $\mathrm{\AA}$/min, and were cooled under plasma at 10-20 \C/min. 

Growth was monitored \textit{in situ} using reflection high energy electron diffraction (RHEED). Post-growth, x-ray photoelectron spectroscopy (XPS) was done via UHV transfer to a PHI 5400 XPS system with a monochromatic Al K$\alpha$ source using 17.9 eV pass energy and an electron flood gun neutralizing source. Binding energies were calibrated by aligning the primary O 1s peak to 530 eV \cite{thapa2021probing}. \textit{Ex situ} x-ray diffraction was done on the films using a Rigaku Smartlab with Cu-K$\alpha$ anode, Ge(220) two-bounce monochromator, and 2D detector. The film thickness was determined using x-ray reflectivity (XRR) and reciprocal space mapping (RSM) was used to determine in-plane epitaxial strain in relation to the substrate. Transport measurements were done using DC van der Pauw method in a Quantum Design Dynacool system. 


High angle annular dark field (HAADF) images of a 10 nm thick SIO sample grown on SrTiO$_3$ were acquired using a Thermo Fisher Themis Z scanning transmission electron microscopy (STEM) operated at 300 kV. Cross sectional samples for STEM were prepared using focused ion beam, with the initial milling at 20 KeV and then subsequently at 5 KeV. The samples were further cleaned using Fischione 1040 Nanomill with the ion beam energy of 500 eV before the STEM imaging. Ir $L_3$ edge x-ray absorption spectra on a 14 nm thick SIO/STO sample were collected at beamline 20-BM of the Advanced Photon Source at Argonne National Laboratory. The x-ray beam was focused by Pt/alumina bilayer coated toroidal mirror and monochromated by a pair of Si(111) crystals. The x-ray polarization of was perpendicular to the film surface normal, and an angle of incidence less than ten degrees was used. Spectra were collected while the sample was spun about the surface normal direction in oreder to mitigate Bragg scattering. A seven element Ge solid state detector was used to collect the Ir $L\alpha$ fluorescence emission as the incident energy was changed.



\begin{figure}[ht]
    \centering
    \includegraphics[width=0.9\linewidth]{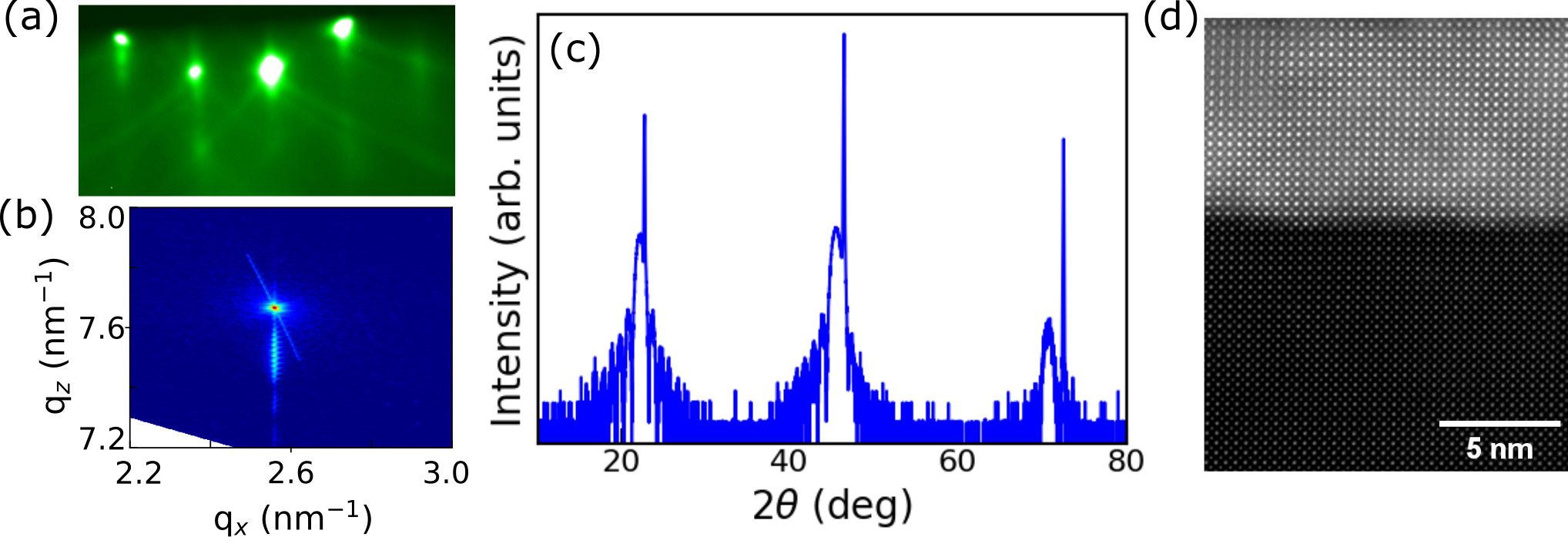}
    \caption{ Structural behavior of strained SrIrO$_{3}$ films grown on SrTiO$_3$. (a) RHEED (b) RSM for (103). (c) XRD (d) Cross-sectional STEM. }
    \label{fig1}
\end{figure}

In Figure \ref{fig1}, we show that high-quality epitaxial SIO films can be grown using metalorganic MBE. RHEED in Figure \ref{fig1}(a) shows that the growth of SIO follows in layered fashion, with the film commensurate with the underlying substrate lattice. SIO has orthorhombic phase in bulk with lattice constants a = 5.60 $\mathrm{\AA}$, b = 5.58 $\mathrm{\AA}$ and  c = 7.89 $\mathrm{\AA}$ \cite{longo_structure_1971}, and the pseudocubic lattice constant for the perovskite form is 3.953 $\mathrm{\AA}$. When grown on closely lattice-matched substrates, the film will be coherently strained, which is verified using RHEED and RSM as shown in Figure \ref{fig1}(a,b). XRD in Figure \ref{fig1}(c) shows that SIO grows as a single phase film. STEM shows the good quality, sharp interface, and commensurate epitaxy of the SIO films in Figure \ref{fig1}(d). XAS for Ir L$_3$ edge is shown in Supplemental Figure S1 and confirms the Ir$^{4+}$ state \cite{liu_synthesis_2017,chaurasia_low-temperature_2021}.

\begin{figure}[ht]
    \centering
    \includegraphics[width=1\linewidth]{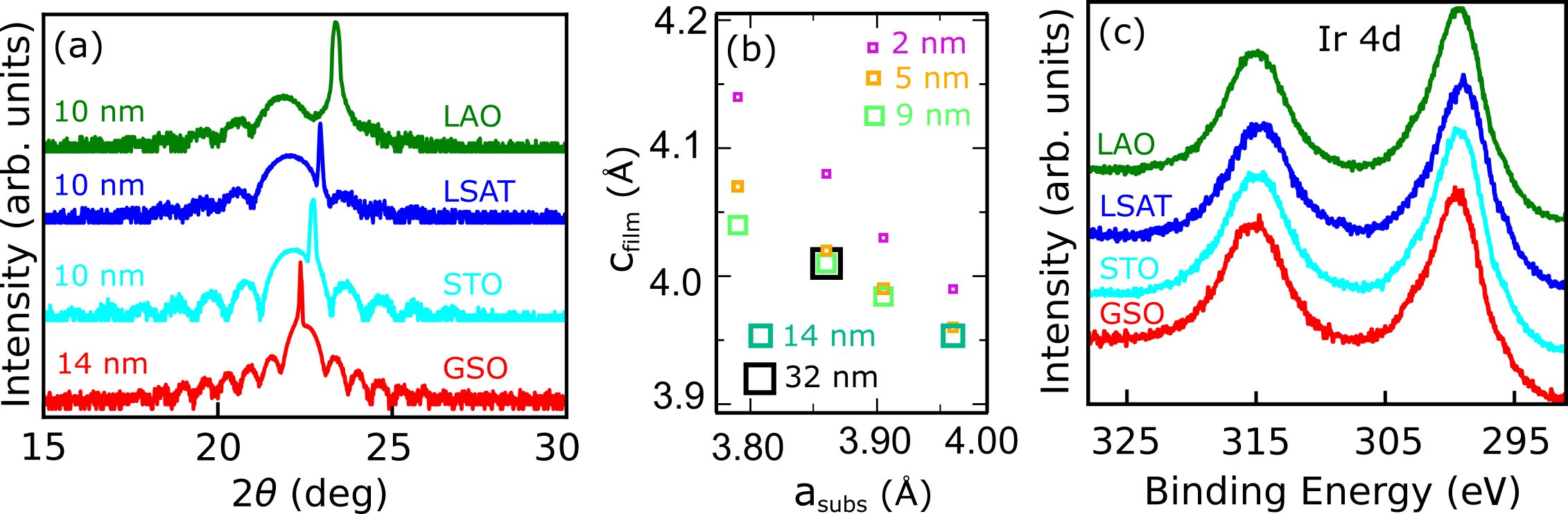}
    \caption{ Comparison of SIO films on different substrates. (a) XRD for (001) peaks. (b) The variation of c-axis lattice constant with the substrate and the thickness. (c) Ir 4d XPS spectra for 2 nm thick films. }
    \label{fig2}
\end{figure}

Figure \ref{fig2} shows the role of epitaxial strain on the SIO films. In Figure \ref{fig2}(a), we compare XRD for the strained SIO films. GSO has a slight tensile strain ($\sim 0.2\%$) while STO, LSAT and LAO result in compressive strains of 1.2\%, 2.4\% and 4.3 \%, respectively. As the lattice constant of the substrates decrease from GSO to STO to LSAT to LAO, we observe an expansion in the c-axis lattice constant for films of same thickness, as shown in Figure \ref{fig2}(b). Strain relaxation will occur if the films are grown sufficiently thick, and only the thinnest film (2 nm) grown on LAO was strained, while only the thickest (32 nm) film on LSAT showed relaxation. Since the lattice mismatch with LAO is large compared to LSAT, it is expected that relaxation will occur sooner for growth on LAO compared to better lattice-matched substrates, which is verified using RHEED and RSM as shown in the supplemental Figure S2. As expected, we also observe that for the most highly strained films (on LAO), the relaxation occurs more quickly with thickness compared to films on the better-matched substrates. Strain will also lead to distortions and expansion in the c-axis lattice parameter which is exhibited in Figure \ref{fig2}(b) where the film XRD peaks shift relative to the substrate peaks. Laue oscillations also confirm the excellent crystallinity and good layering in these films, and help determine the film thickness. XPS measurements of the Ir 4d core level show that the Ir electronic state do not change with strain, as shown in Figure \ref{fig2}(c).

\begin{figure}[ht]
    \centering
    \includegraphics[width=1\linewidth]{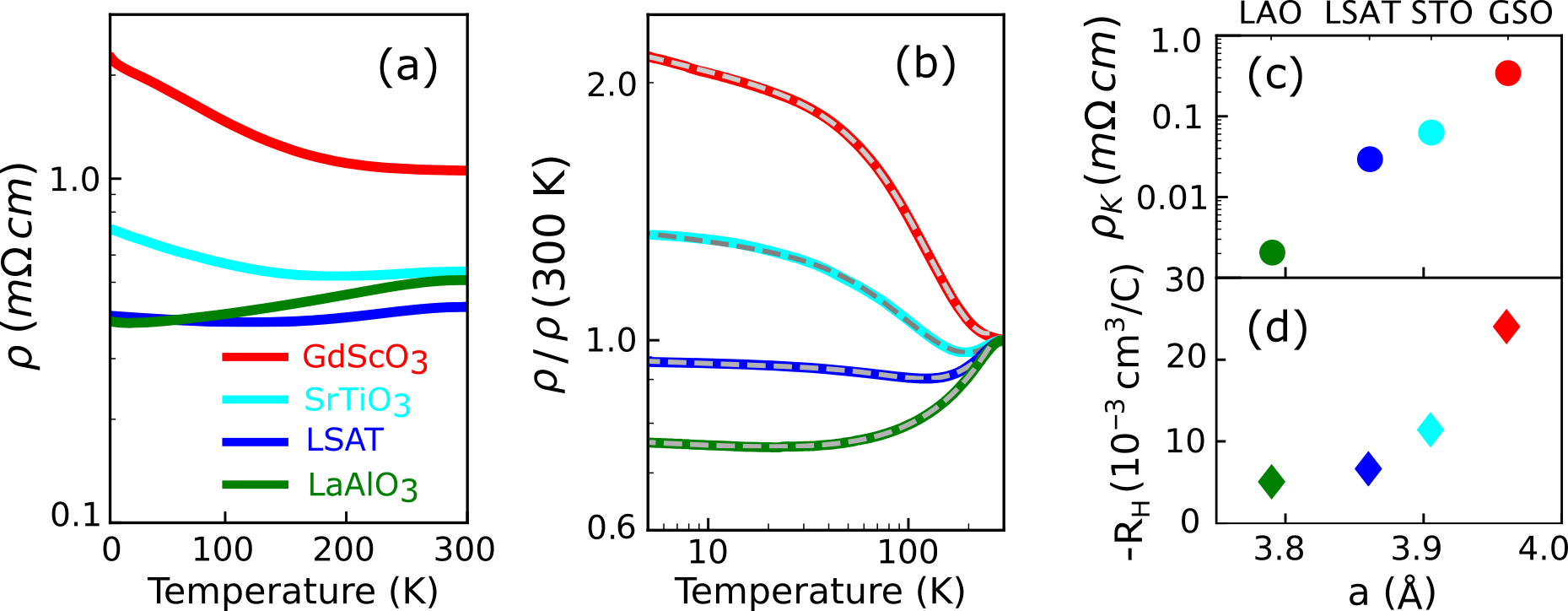}
    \caption{ Transport behavior of 5 nm thick SrIrO$_3$ films. Temperature dependence of (a) resistivity, and (b) resistivity normalized to room temperature value. The dashed lines in (b) show fits to the transport model defined in Equation \ref{fit_eqn}. (c) The zero temperature Kondo resistivity. (d) Hall coefficient at 5 K. }
    \label{fig3}
\end{figure}

Figure \ref{fig3} shows the electronic transport behavior of strained 5 nm thick SIO films. In bulk SIO single crystals, the low temperature resistivity is about 2.2 $\mu\Omega$cm \cite{cao_non-fermi-liquid_2007} and the resistivity of our films is comparable to previously reported values on films \cite{biswas_metal_2014,liu_tuning_2013,nie_interplay_2015,choudhary_semi-metallic_2022}. Both 300 K and 5 K resistivities are comparable to the lowest on record for MBE-grown films using both metalorganic MBE \cite{choudhary_semi-metallic_2022} and electron-beam-assisted MBE \cite{nie_interplay_2015}, indicating the high crystalline quality and low extrinsic impurity concentrations of our films. It can be observed in Figure \ref{fig3}(a) that the least strained film (on GSO) shows an insulating behavior whereas higher compressive strain leads to more metallic behavior. Higher compressive strain was also found to drive a more metallic behavior in SrIrO$_3$/SrTiO$_3$ superlattices \cite{yang_strain-modulated_2020}. However, unlike previous case of strained SrIrO$_3$ on GdScO$_3$ \cite{biswas_metal_2014}, all our films on GSO (up to 14 nm thick) show an insulating behavior but have lower resistivities at 300 K. First principles calculations showed that tensile strain in SIO, which creates a longer bond distance, results in higher carrier hopping energies \cite{kim_electronic_2014}, whereas the reverse can be expected in the case of compressive strain, which qualitatively explains the behavior. The metallic films show an upturn at lower temperatures which may be attributed to Kondo and weak antilocalization effects. Similar upturn in resistivity has been observed in other SIO films \cite{liu_tuning_2013,groenendijk_spin-orbit_2017,choudhary_semi-metallic_2022} but not much emphasis has been placed in understanding the origins or implications of this behavior. The general trend of $\rho$ vs T shows decreasing upturn temperature with increasing strain.

To understand the transport behavior, we modeled the resistivity with contributions from different scattering mechanisms. The temperature dependence of resistivity, $\rho(T)$, can be modeled using contributions\cite{Lee2011,cai_weak_2021,choudhary_semi-metallic_2022} from impurity $\rho_0$, quasiparticle $\rho_1$,  and Kondo scattering $\rho_K$, with added contribution from weak antilocalization $\rho_3$, and given as 

\begin{equation}
    \rho(T) = \rho_0 + \rho_1 \, T^{1.4} + \rho_K \, \left( \frac{T_K'^2}{T^2+T_K'^2} \right)^s - \rho_3 \, \log(T)
    \label{fit_eqn}
\end{equation}

where $T_K' = \sqrt{T_K/(2^{1/s} - 1)}$, s = 0.225 and $T_K$ is the Kondo temperature. This model fits our data well, as shown in Figure \ref{fig3}(b). The strain-dependence of the Kondo contribution to resistivity, as shown in Figure \ref{fig3}(c), suggests the intricate role of strain in the Kondo scattering of charge carriers. The Kondo resistivity, which scales with the concentration of localized magnetic states \cite{Kondo1964}, shows that the strain is responsible for the localized magnetic states that form in the system. Furthermore, as shown in Figure \ref{fig3}(d) the Hall coefficient decreases with increasing compressive strain, which suggests that the net carrier density increases at higher compressive strain. 

We now discuss the nature of the local magnetic states in these films. Unlike in the case of CaCu$_3$Ir$_4$O$_{12}$, wherein incipient Kondo behavior is present due to hybridization between localized Cu and itinerant Ir states \cite{cheng_possible_2013}, there are no clear localized magnetic states available in SIO. Extrinsic impurities are minimal in these films, as indicated by XPS and comparisons to past MBE-grown films \cite{nie_interplay_2015,choudhary_semi-metallic_2022}. Since the amount of these impurities, if any, should be similar for films of the same thickness on different substrates, the localized states must be intrinsic and arise due to the strain. Oxygen vacancies have been predicted to drive ferromagnetism and Kondo behavior in SrTiO$_3$ \cite{lin_consequences_2014,cai_weak_2021} and have also been proposed as the cause of Kondo behavior of SrIrO$_3$ films grown using metalorganic MBE \cite{choudhary_semi-metallic_2022}. Thus, we propose that oxygen vacancies are the source of the local magnetic states and are largely responsible for the Kondo behavior. This prompts a question of what the role of these localized states are in the competition between different energy scales. 

Insulating antiferromagnetic ground state was observed in non-stoichiometric SIO single crystals \cite{zheng_simultaneous_2016} and Sn-substituted SIO \cite{cui_slater_2016} and mechanisms such as carrier doping and chemical pressures were proposed to drive this transition. Density functional theory calculations  have shown that strain has a large role on the oxygen vacancy formation energy \cite{aschauer_strain-controlled_2013}, with tensile strain leading to higher vacancy formation. Similarly, DFT showed the role of epitaxial strain and oxygen vacancies in Sr$_3$Ir$_2$O$_7$ \cite{kim_magnetic_2017}, in which tensile and compressive strain favored antiferromagnetic couplings in both in plane and out of plane directions, and also found that compressive strain favored higher oxygen vacancies. The observation of antiferromagnetic ground state in compressively strained SrIrO$_3$/SrTiO$_3$ superlattices \cite{hao_two-dimensional_2017,yang_strain-modulated_2020} may also be intimately related to the effect of vacancies. It should also be pointed out that due to the similar energy scales there may be large changes in the electronic and magnetic behavior which is why different reports or preparation methods lead to different film behaviors.


Another comment on the transport behavior is on the non-Fermi-liquid (NFL) contribution to the resistivity. Ferromagnetic instability close to the NFL state was found in bulk single crystals of SIO, which lies close to a quantum critical point \cite{cao_non-fermi-liquid_2007}. Cui et al \cite{cui_slater_2016} found that Sn-substituted SrIrO$_3$ shows antiferromagnetic order and gap opening below the transition temperature, and antiferromagnetic order was also found for non-stoichiometric SIO \cite{zheng_simultaneous_2016}. Transport and STM studies also revealed a small gap around 60 meV in SIO confined between SrTiO$_3$ \cite{groenendijk_spin-orbit_2017}, which was attributed to antiferromagnetic order through DFT calculations. Antiferromagnetic order has also been demonstrated using X-ray magnetic scattering in SIO/SrTiO$_3$ superlattices \cite{yang_strain-modulated_2020,gong_reconciling_2022}. The observation of Kondo behavior in our films may thus be a precursor to a magnetic state in strained SIO films.

\begin{figure}[ht]
    \centering
    \includegraphics[width=0.8\linewidth]{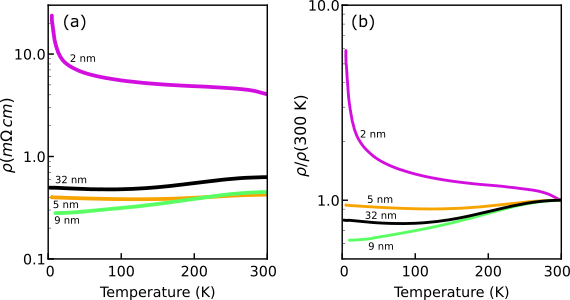}
    \caption{ (a) Resistivity vs. temperature for SIO films on LSAT. (b) The same data normalized to room temperature resistivity to emphasize temperature dependence behavior. }
    \label{fig4}
\end{figure}

We also investigate the role of thickness and strain relaxation in a series of films grown on LSAT substrates, as shown in Figure \ref{fig4}. The film with the lowest thickness shows a fully insulating behavior whereas other films exhibit metallic behavior. Except for the 10 nm thick film the other films show an upturn when cooled and the thickest film has higher resistivity compared to the next thickest film. Among the thinner films, surface scatterings or increased oxygen vacancy concentrations near the film surface are also likely to result in higher resistivity and Kondo effects, thereby leading to the lowest resistivity in the 9 nm thick film. Further studies should help disentangle the role of strain relaxation, oxygen vacancies, and localized moments that drives the metal-to-insulator behavior in SIO, and the possible magnetic states that may emerge in strained films. Density functional theory and dynamical mean field theory modeling of the effects of oxygen vacancies on carrier behavior in SIO would also be worthwhile. 

In summary, we show a systematic metal-to-insulator transition in epitaxially strained SrIrO$_3$ films. Higher compressive strain is found to show metallic behavior while lower compressive strain and even a slight tensile strain results in insulating states. The observation of Kondo behavior in resistivity shows the presence of localized magnetic states associated with strain, which likely arise due to oxygen vacancies. Our results suggest proximity of the strained $n=\infty$ perovskite iridate to a magnetic ground state and a quantum critical point that can be tuned via strain and oxygen vacancy concentrations. Further work could explore alternative avenues to electron-dope SrIrO$_3$ to examine how tuning of electron concentrations can be used to control the strength of Kondo scattering and the emergence of a Kondo insulating state in the material.


\section*{Acknowledgements}
We thank Bharat Jalan for helpful discussions. Research at Auburn University was supported by the U.S. Department of Energy (DOE), Office of Science, Basic Energy Sciences (BES), under Award DE-SC0023478 (film synthesis, X-ray diffraction, and electrical transport, G.R. and R.B.C.) and by the National Science Foundation (NSF) under Award DMR-2045993 (X-ray photoelectron spectroscopy, T.T.). X-ray diffraction measurements were performed at Auburn on a system supported by the NSF Major Research Instrumentation (MRI) program under award number DMR-2018794. G.C.O and J.H acknowledge funding support from NSF under Award DMR-1847964 (electron microscopy). This research used resources from the Advanced Photon Source, an Office of Science User Facility operated for the U.S. Department of Energy (DOE) Office of Science by Argonne National Laboratory and was supported by the U.S. DOE under Contract No. DE-AC02-06CH11357.

\bibliography{main}

\end{document}


\title{Strain-dependent Insulating State in epitaxial SrIrO$_{3}$ Films\\Supplemental Information}

\author{Gaurab Rimal}
\email{gaurab.rimal@wmich.edu}
\affiliation{Department of Physics, Auburn University, Auburn, AL, USA}
\affiliation{Department of Physics, Western Michigan University, Kalamazoo, MI, USA}
\author{Tanzila Tasnim}
\affiliation{Department of Physics, Auburn University, Auburn, AL, USA}

\author{Gabriel Calderon Ortiz}
\affiliation{Department of Materials Science and Engineering, Ohio State University, Columbus, OH, USA}
\author{George E. Sterbinsky}
\affiliation{Advanced Photon Source, Argonne National Laboratory, Lemont, IL, USA}
\author{Jinwoo Hwang}
\affiliation{Department of Materials Science and Engineering, Ohio State University, Columbus, OH, USA}
\author{Ryan B. Comes}
\email{ryan.comes@auburn.edu}
\affiliation{Department of Physics, Auburn University, Auburn, AL, USA}

\newcommand{\red}[1]{\textcolor{red}{#1}}
\newcommand{\blue}[1]{\textcolor{blue}{#1}}
\newcommand{\green}[1]{\textcolor{green}{#1}}

\maketitle

\section{Substrate dependence of RSM}

\begin{figure}[ht]
    \centering
    \includegraphics[width=0.9\linewidth]{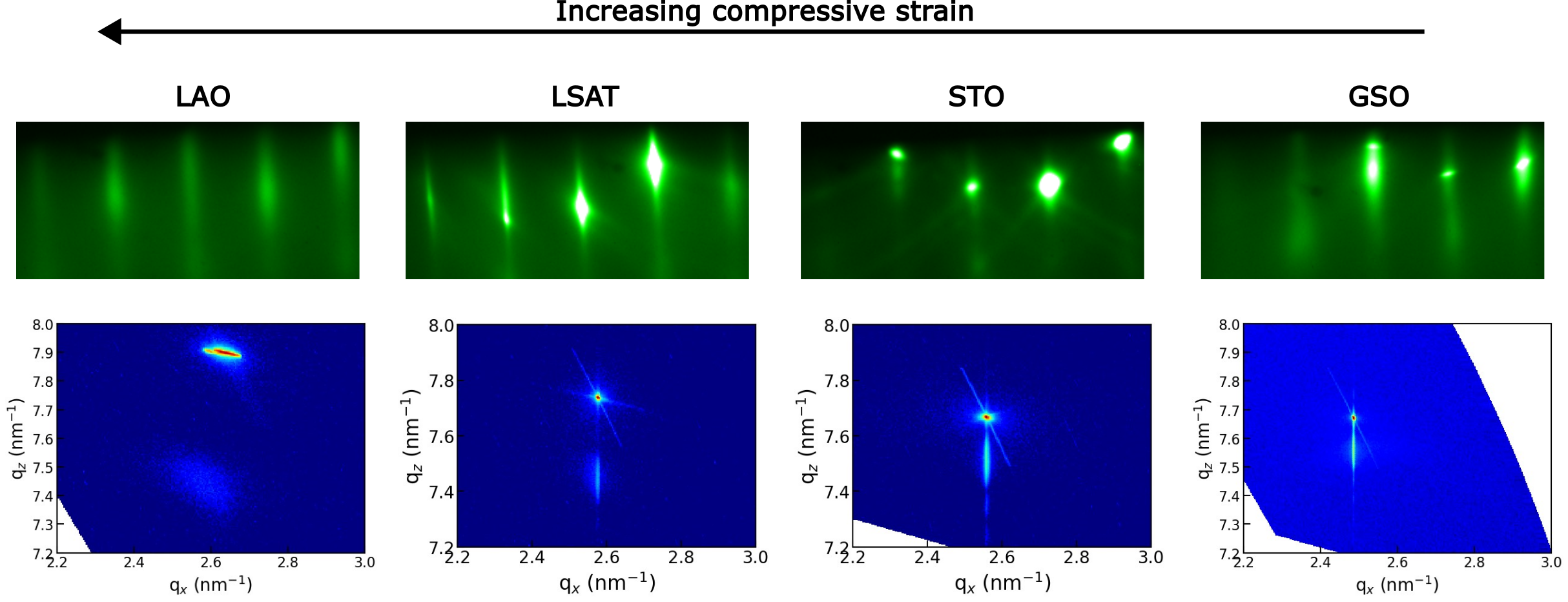}
    \caption{RHEED and RSM for 10 nm thick films }
    \label{fig1}
\end{figure}

\newpage

\section{Scanning transmission electron microscopy}
\begin{figure}[ht]
    \centering
    \includegraphics{STEM EDS.png}
    \caption{(a) HAADF STEM image of SrIrO$_3$/SrTiO$_3$ sample; (b-e) Energy dispersive X-ray spectroscopy (EDS) map of (b) Ti, (c) Ir, (d) Sr, and (e) O. }
    \label{fig2}
\end{figure}
\begin{figure}[ht]
    \centering
    \includegraphics{STEM EDS Comp.png}
    \caption{(a) HAADF STEM image of SrIrO$_3$/SrTiO$_3$ sample from Fig 3 showing line profile of EDS composition; (b) EDS composition with depth in the sample.}
    \label{fig3}
\end{figure}

\newpage
\section{XAS}

\begin{figure}[ht]
    \centering
    \includegraphics[width=0.5\linewidth]{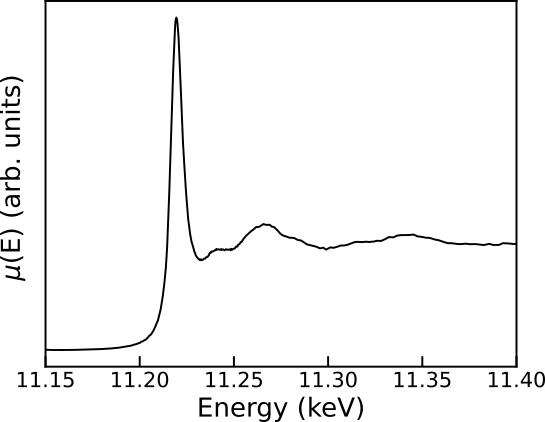}
    \caption{ Ir L$_3$ edge XAS with in-plane polarization. }
    \label{fig4}
\end{figure}